\documentclass[twocolumn]{aastex63}
\usepackage{amsmath}
\received{XXX}
\revised{XXX}
\accepted{XXX}
\submitjournal{AJ}

\shorttitle{}
\shortauthors{Steinwandel et al.}
\begin{document}

%\title{An analytic prescription on how disc galaxies obtain their magnetic field and simultaneously drive outflows \footnote{Released on June, 10th, 2019}}
\title{Driving galactic outflows with magnetic fields at low and high redshift\footnote{Released on August, 13th, 2020}}

\correspondingauthor{Ulrich P. Steinwandel}
\email{usteinw@usm.lmu.de, uli@mpa-garching.mpg.de}

\author[0000-0001-8867-5026]{Ulrich P. Steinwandel}
\affiliation{University Observatory Munich, Scheinerstr. 1, D-81679 Munich, Germany}
\affiliation{Max Planck Institute for Astrophysics, Karl-Scharzschildstr. 1, D-85748, Garching, Germany}

\author{Klaus Dolag}
\affiliation{University Observatory Munich, Scheinerstr. 1, D-81679 Munich, Germany}
\affiliation{Max Planck Institute for Astrophysics, Karl-Scharzschildstr. 1, D-85748, Garching, Germany}

\author{Harald Lesch}
\affiliation{University Observatory Munich, Scheinerstr. 1, D-81679 Munich, Germany}

\author{Andreas Burkert}
\affiliation{University Observatory Munich, Scheinerstr. 1, D-81679 Munich, Germany}
\affiliation{Max Planck Institute for Extraterrestrial Physics, Giessenbachstr. 1, D-85748, Garching, Germany}

%% Note that the \and command from previous versions of AASTeX is now
%% depreciated in this version as it is no longer necessary. AASTeX 
%% automatically takes care of all commas and "and"s between authors names.

%% AASTeX 6.3 has the new \collaboration and \nocollaboration commands to
%% provide the collaboration status of a group of authors. These commands 
%% can be used either before or after the list of corresponding authors. The
%% argument for \collaboration is the collaboration identifier. Authors are
%% encouraged to surround collaboration identifiers with ()s. The 
%% \nocollaboration command takes no argument and exists to indicate that
%% the nearby authors are not part of surrounding collaborations.

%% Mark off the abstract in the ``abstract'' environment. 
\begin{abstract}
Although playing a key role for our understanding of the evolution of galaxies, the exact way how observed galactic outflows are driven is still far from being understood and therefore our understanding of associated feedback mechanisms that control the evolution of galaxies is still plagued by many enigmas.
In this work we present a simple toy model that can provide insight on how non-axis-symmetric instabilities in galaxies (bars, spiral-arms, warps) can lead to local exponential magnetic field growth by a radial flows beyond the equipartition value by at least two orders of magnitude on a time-scale of a few $100$ Myr. Our predictions show that the process can lead to galactic outflows in barred spiral galaxies with a mass loading factor $\eta \approx 0.1$, in agreement with our numerical simulations. Moreover, our outflow mechanism could contribute to an understanding of the large fraction of bared spiral galaxies that show signs of galactic outflows in the \textsc{chang-es} survey. 
Extending our model shows the importance of such processes in high redshift galaxies by assuming equipartition between magnetic energy and turbulent energy. Simple estimates for the star formation rate (SFR) in our model together with cross-correlated masses from the star-forming main-sequence at redshifts $z\sim2$ allow us to estimate the outflow rate and mass loading factors by non-axis-symmetric instabilities and a subsequent radial inflow dynamo, giving mass loading factors of $\eta \approx 0.1$ for galaxies in the range of M$_{\star}=10^9 - 10^{12}$ M$_{\odot}$, in good agreement with recent results of \textsc{sinfoni} and \textsc{kmos}$^{3\mathrm{D}}$.

\end{abstract}

%% Keywords should appear after the \end{abstract} command. 
%% See the online documentation for the full list of available subject
%% keywords and the rules for their use.

%% From the front matter, we move on to the body of the paper.
%% Sections are demarcated by \section and \subsection, respectively.
%% Observe the use of the LaTeX \label
%% command after the \subsection to give a symbolic KEY to the
%% subsection for cross-referencing in a \ref command.
%% You can use LaTeX's \ref and \label commands to keep track of
%% cross-references to sections, equations, tables, and figures.
%% That way, if you change the order of any elements, LaTeX will
%% automatically renumber them.
%%
%% We recommend that authors also use the natbib \citep
%% and \citet commands to identify citations.  The citations are
%% tied to the reference list via symbolic KEYs. The KEY corresponds
%% to the KEY in the \bibitem in the reference list below. 

\section{Introduction} \label{sec:intro}

Observations in the radio continuum usually indicate a radially declining magnetic field at around $10$ $\mu$G, observed in a wide range of local spiral galaxies. Locally the ordered and the turbulent component show different scaling between spiral- and inter-arm regions, where the ordered magnetic field is observed to be higher in the inter-arm regions compared to the spiral arms \citep[e.g.][and references therein]{Beck2015}. Furthermore, recent observations indicate that many of these galaxies show signs of galactic outflows \citep[e.g.][]{Krause2018, Krause2020, Miskolczi2019, Stein2019, Schmidt2019, Mora2019}. On top of that there is reported H-$\alpha$ emission from nearby galaxies \citep{Vargas2019} in magnetically active edge-on galaxies, further indicating non-star forming gas. \\
Therefore, observational it is well constrained that a lot of local magnetised spiral galaxies appear to be quite active in terms of their outflow activity. We suggest that the presence of the magnetic field can self-consistently launch these outflow and account for the observed X-shaped structure in the halo-field \citep{Golla1994, Tuellmann2000, Krause2006, Heesen2009, Soida2011, Stein2019} due to a wind that is launched by the magnetic pressure. On top of self-consistently generating the observed field structure such a magnetic driven process can also directly account for the observed strong fields in the halos of galaxies of around 10$^{-7}$ G, that would then be amplified in the galactic disc due to dynamo action and transported to the halo by the magnetic wind. The process that we suggest consists out of the following steps to launch the outflow:
\begin{enumerate}
 \renewcommand{\labelenumi}{\roman{enumi}}
    \item Amplifying the field to equipartition strength via the small-scale turbulent dynamo.
    \item Ordering the field with the $\alpha$-$\Omega$-dynamo on the large-scales.
    \item Generating a super-equipartition regime with a low plasmabeta by radial inflows dud to gravitational instabilities of the galactic disc.
    \item Generating an open field geometry to launch the outflow.
\end{enumerate}
Observational studies suggest that $\sim \mu$G magnetic fields in galaxies are in agreement with large-scale dynamo action ($\alpha$-$\Omega$-dynamo) in the galactic disc. This argumentation suggests that the large-scale dynamo amplifies a weak magnetic seed field up to the equipartition (a few $\mu$G) by small-scale vertical motion of buoyant (supernova) heated bubbles that are lifted up and get sheared by the Coriolis force ($\alpha$-effect). In rotating spiral galaxies the magnetic field lines are then supposed to be twisted and folded by the large-scale rotation of the disc ($\Omega$-effect). While this picture of the large-scale dynamo is a good model to explain the magnetic field structure in an already evolved spiral galaxy, it represents an oversimplification of how magnetic fields are amplified in the Universe.\\
First, the amplification time-scale of the $\alpha$-$\Omega$-dynamo in combination with the tiny primordial seed fields in order of $10^{-20}$ G \citep[e.g.][]{Biermann1950, Harrison1970} cannot explain the observed $\mu$G fields today, even if one assumes that the Milky Way (MW) formed $13.8$ Gyr ago as a fully developed disc, which is in strong disagreement with the findings of large-volume simulations of the Universe \citep[e.g.][]{Teklu2015, DeFelippis2017, Lagos2017, Zjupa2017}. In addition, observations indicate that galaxies have already very strong magnetic fields at high redshift that are at least as high as the magnetic field today \citep{Perry1993, Bernet2008, Kronberg2008, Wolfe2008} which furthermore strengthens the timescale argument (see discussion in section \ref{sec:alpha}).\\
Second, the galactic magnetic field is observed to have a quadrupolar structure. Theoretical models for the $\alpha$-$\Omega$-dynamo favour the growth of the dipol mode (see discussion in section \ref{sec:alpha}) and thus the $\alpha$-$\Omega$ dynamo as the main amplification process is in tension with the observed field structure.
Third, the theoretical model for the $\alpha$-$\Omega$ dynamo has boundary conditions that would lead to an excess flux, inconsistent with observed outflow rates in galaxies (see discussion in section \ref{sec:alpha}).\\
However, from the theoretical point of view this can beautifully be resolved by considering the small-scale-turbulent dynamo \citep[e.g.][]{Kraichnan1967, Kazantsev1968, Zeldovich1983, Kazantsev1985, Kulsrud1992, Kulsrud1997, Xu2020} which amplifies magnetic fields due to stretching, twisting and subsequent folding of field lines by turbulence driven in the interstellar medium (ISM). This can either happen due to large-scale accretion flows or stellar feedback shown in various numerical simulations \citep[e.g.][]{Beck2012, Butsky2017, Hopkins2020a, Martin2018, Martin2020, Pakmor2013, Pakmor2017, Rieder2016, Rieder2017a, Rieder2017b, Steinwandel2019, Steinwandel2020, Su2018} This dynamo operates on time-scales of a few $10$ Myrs and can quickly amplify a weak primordial seed field, at the highest redshifts. Hereby, small-scale turbulence is the main driver of magnetic field amplification in the ISM and removes the constraint of any ordered large-scale motions of disc galaxies at high redshift. This is in complete agreement with the recent results of galaxy scale simulations and cosmological zoom-in simulations of MW-like galaxies (see section \ref{sec:sims} for a detailed discussion). In this scenario the purpose of the $\alpha$-$\Omega$ dynamo is to retain the field once the small-scale turbulent dynamo has generated it, before the field vanishes due to magnetic dissipation.\\
Furthermore, a magnetic field, established in this fashion contributes massively to the energy density and could potentially trigger a large-scale outflow in a galaxy if there is a mechanism that can efficiently amplify the magnetic field beyond the equipartition field strength. In the following we will show that this can be achieved by bar formation and radial inflows that will drive the amplification of the field, while the Parker-Instability provides the field geometry necessary for outflow launching. We therefore suggest that the problem of magnetic field amplification and the cause for galactic outflows are highly connected problems as the magnetic field can contribute a significant amount to the midplane pressure in the ISM.\\
In the following we will develop a framework that will explain the formation of a magnetic driven outflows due to a fast track dynamo caused by a non-axis-symmetric perturbation (bar, spiral arm, warp) in the galactic disc based on the assumption that the small-scale turbulent dynamo amplified the magnetic field beforehand to equipartition field strength and the $\alpha$-$\Omega$-dynamo generates the large-scale field structure.
Therefore, we first discuss magnetic field amplification in a galactic context in section \ref{sec:analytic} and show how non-axis symmetric instabilities can exponentially amplify the magnetic field strength and predict the outflow-rate based on magneto-centrifugal wind theory. In section \ref{sec:sims} we show that these simple estimates are consistent with results that can be obtained with numerical simulations at $z \sim 0$. In section \ref{sec:highz} we explain how our derived model can impact galactic outflows at $z \sim 2$. In section \ref{sec:conclusions} we summarise our results.    

\section{Theory of magnetic field amplification} \label{sec:analytic}

\subsection{Magnetic seed fields\label{sec:seed}}

Generally, it is assumed that magnetic fields originate from tiny seed-fields that arise in the early universe. In the following we will briefly summarise the various processes for seed-field generation.\\ The Biermann-battery \citep{Biermann1950} is the most popular process to generate primordial magnetic seeds. It is initiated by non-linear terms in Ohm's law which lead to a source term in the induction equation that is proportional to $\nabla p  \times \nabla \rho$. Thus, a tiny magnetic field is induced when the gradients of pressure and density are miss-aligned. This yields a seed field well below $10^{-21}$ G. However, ionisation fronts during the epoch of reionisation \citep[EoR, e.g.][]{Spergel2007} could lead to a more efficient Biermann-battery process that sets an upper limit of $~10^{-17}$ \citep[e.g.][]{Gnedin2000}. \\
\citet{Harrison1970} argues that the rotating motion in a sphere of plasma is decoupled from the radiation field at high redshift, due to the increased photon mass a that time. Due to Thompson scattering with the photons, the electrons slow down and induce a current. This induces a magnetic field. As this magnetic seed field increases, it induces an electric field that stabilises the rotation of the electrons in the gas sphere. A similar mechanism is proposed by \citet{Matarese2004}. Both mechanisms result in a seed field way below $10^{-20}$ G.\\
\citet{Demozzi2010} point out that a tiny seed field of the order of $10^{-32}$ could be generated on Mpc scales during inflation.\\
However, there are other mechanisms suggested to generate even higher magnetic seed fields for example by the seeding of supernovae. Seeding the magnetic field by supernovae could generate a background field of up to $10^{-9}$ G in the Galaxy following the studies of \citet{Rees1987, Rees1994, Rees2005, Rees2006}. The idea behind this approach is that the magnetic field is generated during stellar evolution (e.g. due to an $\alpha$-$\Omega$-dynamo) and is distributed to the ISM when the star ends its life in a supernova. This can be used to estimate the magnetic field strength released over the galactic lifetime of roughly $10$ Gyr\footnote{This refers to the lifetime of the Galaxy as a fully developed disc.} following \citet{Beck2012} who estimate 10$^{-9}$ G Gyr$^{-1}$ for a total supernova rate of $10^8$ within the volume of the Milky Way which is roughly $300$ kpc$^{3}$. \\ 
Finally, some authors argue that one can generate very strong sub-equipartition field of around $10^{-7}$ G due to plasma instabilities like the Weibel-Instability \citep[e.g.][]{Schlickeiser2003, Lazar2009}. This is a very intriguing picture because it basically solves the magnetic seed problem alongside with the amplification problem by providing seeds that are just one order of magnitude below the equiaprtition value of the magnetic field in nearby galaxies. However, \citet{Schlickeiser2003} point out that growth only occurs for very high Mach-numbers with $\mathcal{M} > 43$. Galaxy-cluster simulations \citep[e.g.][]{Miniati2001, Vazza2011} from various groups show that there are very few shocks with $\mathcal{M} > 43$ and almost none with $\mathcal{M} > 70$ as even pointed out by \citet{Schlickeiser2003}. In combination with the fact that the Weibel-Instability amplifies the magnetic field on very small scales, this renders the question if the instability can generate a coherent high background field on kpc or even Mpc scales.
 
\subsection{\texorpdfstring{Amplification due to the $\alpha$-$\Omega$ dynamo}{sec:alpha}}

%\subsection{Amplification due to the $\alpha$-$\Omega$ dynamo}
\label{sec:alpha}
\citet{Larmor1919} pointed out that strong magnetic fields could be obtained in a dynamo process in stellar bodies and first attempts for cosmic magnetic field amplification were made considering axis-symmetric velocities by splitting the field in its poloidal and toroidal components. It is straightforward to see that toroidal fields can be generated from poloidal fields by differential rotation \citep[e.g.][]{Kulsrud2005}. This can be understood by considering an initial poloidal field in a deferentially rotating disc. The field poloidal field lines will move with different velocities in the differentially rotating frame of a galactic disc and some toroidal field will be generated. Vice versa, if one starts from a purely toroidal field rotating the disc will only keep the symmetry of the system and there is no amplification of the poloidal component. Thus, rotation alone will not amplify the magnetic field as it will only convert a poloidal field component to a toroidal field component. Therefore, it is impossible to amplify a weak axis-symmetric magnetic field by pure axis-symmetric motions\footnote{If this would be valid, this would correspond to purely linear growth and one can easily show that one would need order of $10^{14}$ rotations of the MW to reach this field strength while it could have rotated $~50$ times even if it formed with today's properties at redshift $20$.} \citep[see][]{Cowling1934} to a substantial field strength.\\ 
\citet{Parker1955} pointed out that one could generate a significant poloidal field from an initial toroidal field by introducing rising convection cells\footnote{In disc galaxies rising convection cells could be interpret as supernova-remnants that experience backward motion due to the Coriolis force.} that are twisted by the Coriolis force of a rotating body when combined with differential rotation. The distortion of the poloidal component would inflict growth in the toroidal component and one obtains exponential growth of the form $~e^{\gamma t}$. The \citet{Parker1955} dynamo model can be generalised in mean field dynamo approximation \citep{Steenbeck1966}, where turbulent motions are treated by the kinetic helicity, quantified by $\alpha = -\tau/3 <\mathbf{v} \cdot \nabla \times \mathbf{v}>$. Their mixing can be quantified with the turbulent resistivity $\beta=\tau/2 <\mathbf{v} \cdot \mathbf{v}>$. Introducing fluctuations of velocity and magnetic field of the form $\mathbf{w}=\mathbf{w^{0}} + \mathbf{w'}$, where $\mathbf{w}$ is an arbitrary vector quantity. This yields the dynamo equation in thin disc approximation in cylindrical coordinates:
\begin{align}
    \frac{\partial B_{r}}{\partial t} &= - \frac{\partial }{\partial z} \left(\alpha B_{\phi}\right) + \beta \frac{\partial^2B_{r}}{\partial z^2},\\
    \frac{\partial B_{\phi}}{\partial t} &= -\Omega B_{r} + \beta \frac{\partial^2 B_{\phi}}{\partial z^2}
\end{align}
This can be solved as an Eigenvalue problem with boundary conditions of a thin disc with scale height $h$ where $B_{r}$ and B$_{\phi}$ vanish at $\pm h$ (only valid if $\beta$ is large) in reduced coordinates yielding:
\begin{align}
    \gamma' B'_{r} &= -\frac{\partial (z' B'_{\varphi})}{\partial z'} + \frac{\partial^2 B'_{r}}{\partial z'^{2}}, \\
    \gamma' B'_{\varphi} &= D B'_{r} + \frac{\partial^2 B'_{\varphi}}{\partial z'^{2}}, 
\end{align}
with $z'=z/h$, $t'= \beta t/h^2$, $\gamma'= \gamma h^2/\beta$, $B_{\varphi} = B'_{\varphi}(\beta/h\alpha_{0})$, $B_{r} = B'_r(\beta/h\alpha_{0})$, $\alpha_{0} = \alpha h/z$ and the dimensionless dynamo number $D=-\Omega \alpha_{0} h^3/\beta^2$. The solution shows exponential growth for $D < D_\mathrm{crit}$ where D$_\mathrm{crit} <-4$ gives rise to dipol modes and D$_\mathrm{crit} <-13$ gives rise to quadrupol modes on the time-scale $h^2/\beta$. The growth-time depends on the disc scale height $h$ and the exact value of $\beta$. \\
\citet{Parker1979} and \citet{Ruzmaikin1988} give an estimate of $0.5$ Gyr for a turbulent velocity of $10$ km s$^{-1}$, a supernova injection radius of $100$ pc and a disc scale height of $300$ pc, which seems to agree with observed values in the MW. On this time-scale one can amplify a field of $10^{-14}$ G to $10^{-6}$ over the lifetime of the galactic disc of $6$ Gyr.  However, we already discussed in section \ref{sec:seed} that there are good arguments to assume that primordial seed fields are much lower than $10^{-14}$ G. Therefore, the $\alpha$-$\Omega$-dynamo has a time-scale problem.\\ Furthermore, the magnetic field structure of the Galaxy is observed to be quadrupolar but the $\alpha$-$\Omega$-dynamo favours a dipol structure. Finally, we note that the boundary conditions for the dynamo equations are problematic as well. In ideal MHD the field is locked to the fluid. To remove magnetic field at the edges, interstellar matter must vanish from the galaxy, which leads to problems in both the enrichment history of the halo and its energetics \citep[see][in their chapter 9 for a detailed discussion]{Kulsrud2008}. However, the large time-scale in combination with the reality of small seed fields from cosmology is the biggest problem and it heavily depends on the estimate of $\beta$. \\
We note that this could be resolved with a better approximation for $\beta$ \citep[e.g.][]{Brandenburg1995, Poezd1993} or a modified dynamo model based on super bubbles \citep[e.g.][]{Ferriere1992a, Ferriere1992b, Ferriere1993a, Ferriere1993b, Ferriere1996, Ferriere1998,Ferriere2000}. \\
Furthermore, we note that a seed field generated by local plasma-instabilities could generate higher seed field around $10^{-7}$ G as we pointed out in section \ref{sec:seed}. However, there is one crucial thing to keep in mind with this picture. While the large-scale dynamo could amplify such a strong seed field over the timescale of a few Gyr to euipartition it renders the problem that overdense structures have to act as seed for the Weibel-Instability. While this could generate the magnetic field in galaxy clusters and massive galaxies this formation scenario remains in question because it can intrinsically not explain the strong intergalactic magnetic fields and the magnetic fields in voids \citep[e.g.][]{Durrer2013}. This could be resolved by galactic winds. However, these winds would have to be quite strong to reach higher magnetic field strengths in voids. 

\subsection{Amplification due to the small-scale turbulent dynamo}
\label{sec:turb_dyn}
\begin{figure}
    \centering
    \includegraphics[scale=0.4]{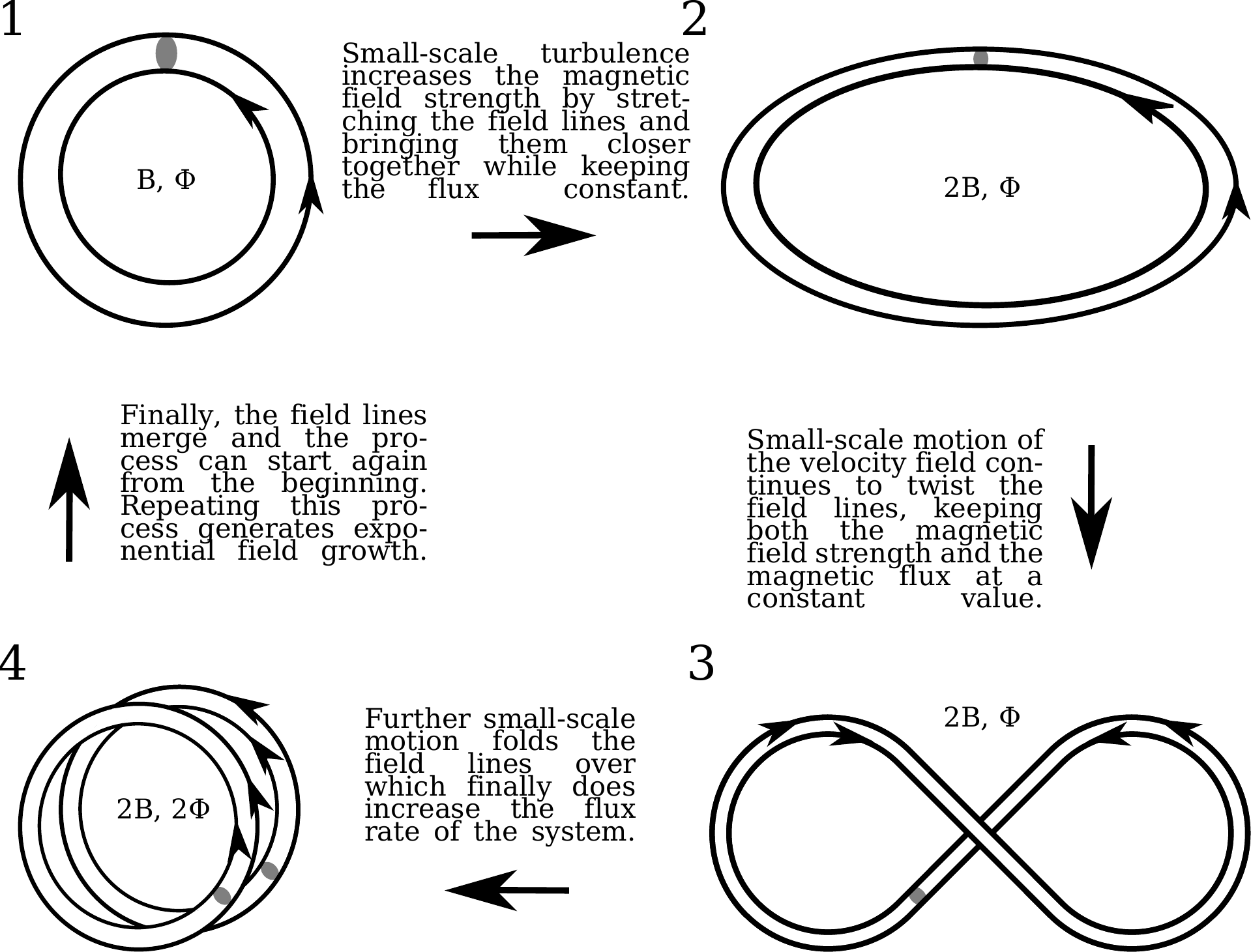}
    \caption{Sketch of the stretching twisting and folding of magnetic field lines in the small-scale turbulent dynamo. In this picture the field strength is increased by stretching a field line at constant magnetic flux. Small-scale turbulent motion then twist and folds the field line which also increases the magnetic flux. Subsequent stretch twist and fold events then lead to exponential growth of the field.}
    \label{fig:twist}
\end{figure}
\begin{figure*}
    \centering
    \includegraphics[scale=0.7]{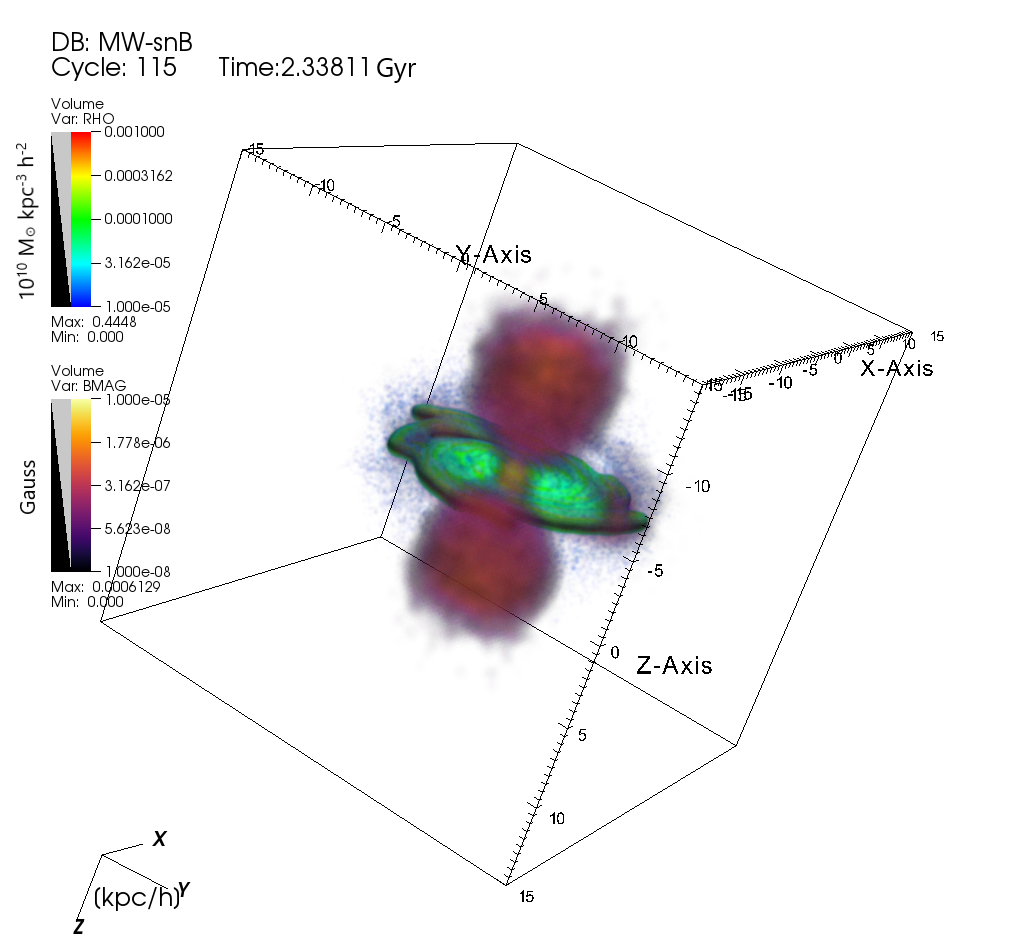}
    \caption{We show a 3d-volume rendering of the gas density (rainbow colours) and the magnetic field-structure (plasma colours) of our simulated MW-analogue. We can clearly see the biconal shape of the outflow in the magnetic field structure. We show the outflow shortly after it has been launched in combination of the bar formation process, the Parker-Instability and adiabatic compression of the magnetic field in the centre of the galaxy.}
    \label{fig:volume}
\end{figure*}
\begin{figure*}
    \centering
    \includegraphics[scale=1.1]{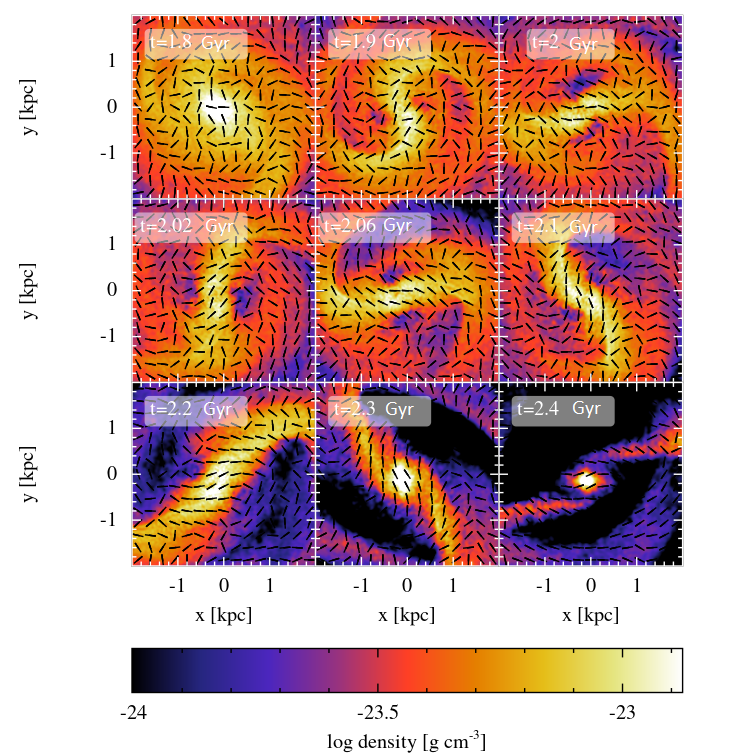}
    \caption{Time evolution of the bar formation process for nine different snapshots. The colour shows he projected gas density and the small arrows indicate the direction of the magnetic field vector. At early times of the bar formation process we can clearly see that the magnetic field lines are completely uncorrelated and have no preferred direction. Once the bar is more prominent, we see that the magnetic field aligns with the bar (from $t=2$ Gyr). From this moment on the mass inflow is heavily supported by the bar, as mass can move force free alongside the magnetic field lines. Further, this leaves the radial component of the magnetic field roughly constant in the centre and the toroidal component is amplified via the radial inflow dynamo which subsequently drives mass flow into the CGM.}
    \label{fig:bars}
\end{figure*}
It has been pointed out that the magnetic field could be generated during the formation process \citep{Pudritz1989, Kulsrud1997} of galaxies and galaxy clusters due to strong turbulence, driven by shocks in the high redshift ISM and ICM. These shocks lead to miss-aligned pressure and density gradients and induce a magnetic field. This leads to a magnetic field growth proportional to the eddy-turnover time of the smallest eddies.\\
This process has been studied extensively in theory \citep{Kraichnan1967,Kazantsev1968, Kulsrud1992, Subramanian2002, Boldyrev2004} and is well understood. Mathematically, the idea is to derive the distribution of the power in the magnetic field under the assumption that velocity and magnetic field can be Fourier decomposed. For random velocities the magnetic power spectrum $P_\mathrm{M}(k)$ is given as the ensemble average of the magnetic energy density:
\begin{align}
    E_\mathrm{mag} = \frac{<\mathbf{B}^2>}{8 \pi} = \int P_\mathrm{M}(k) dk.
    \label{eq:mag_dens}
\end{align}
The evolution of $P_\mathrm{M}(k)$ is given as \citep[e.g.][]{Kulsrud2008}:
\begin{align}
    \frac{\partial P_\mathrm{M}(k)}{\partial t} = \int K(k,k_{0}) M(k_{0}) dk_{0} - 2\beta k^2 P_\mathrm{M}(k),
    \label{eq:mag_struct}
\end{align}
with the structure function $K$ and the turbulent resistivity $\beta$. Combining equations \ref{eq:mag_dens} and \ref{eq:mag_struct} one can find:
\begin{align}
  \frac{dE_\mathrm{mag}}{dt} = 2 \gamma E_\mathrm{mag}, 
\end{align}
which directly implies that the magnetic field strength doubles on the timescale of the eddy-turn-over time. We show this process schematically in Figure \ref{fig:twist}. The idea is that the field lines are stretched, twisted and folded by small-scale turbulence. It is worthwhile to note that such a process needs a three-dimensional approach as the folding of the field lines requires an off plane motion. The growth rate is directly given as the smallest eddy turn-over-time and the energy is transported via an inverse turbulence cascade to the larger scales. In the kinematic regime the evolution of $P_\mathrm{M}(k)$ is given via:
\begin{align*}
    \frac{\partial P_\mathrm{M}(k)}{\partial k} &= \frac{\gamma}{5} \left(k^2\frac{\partial^2 P_\mathrm{M}(k)}{\partial k^2} - 2 k\frac{\partial P_\mathrm{M}(k)}{\partial k} + 6 P_\mathrm{M}(k)\right)
\end{align*}
\begin{align}
    &- 2k^2 \lambda_\mathrm{res} P_\mathrm{M}(k), 
\end{align}
with the resistivity $\lambda_\mathrm{res}$. This can be solved in Fourier space and one obtains: 
\begin{align}
    P_{M}(k,t) \propto e^{3/4 \gamma t} k^{3/2},
\end{align}
yielding exponential growth of Kazantsev modes with $k^{3/2}$. Easy estimates show that this dynamo has eddy turn-over-times that are smaller by a factor of $100$ compared to the free fall time of proto galactic halo. While this can easily lead to field strengths that are larger by a factor of $1000$ compared to observed fields in today's spiral galaxies, the dynamo saturates when equipartition of the magnetic energy and the turbulent velocity of the smallest eddy is reached. In this picture one only needs the $\alpha$-$\Omega$-dynamo at low redshift to explain the large-scale correlation of the field and the field is mainly amplified via the small-scale turbulent dynamo. This results in a so called $\alpha^2$-$\Omega$-dynamo that can generate observed fields in strength and structure. 
\begin{figure}
    \includegraphics[scale=0.5]{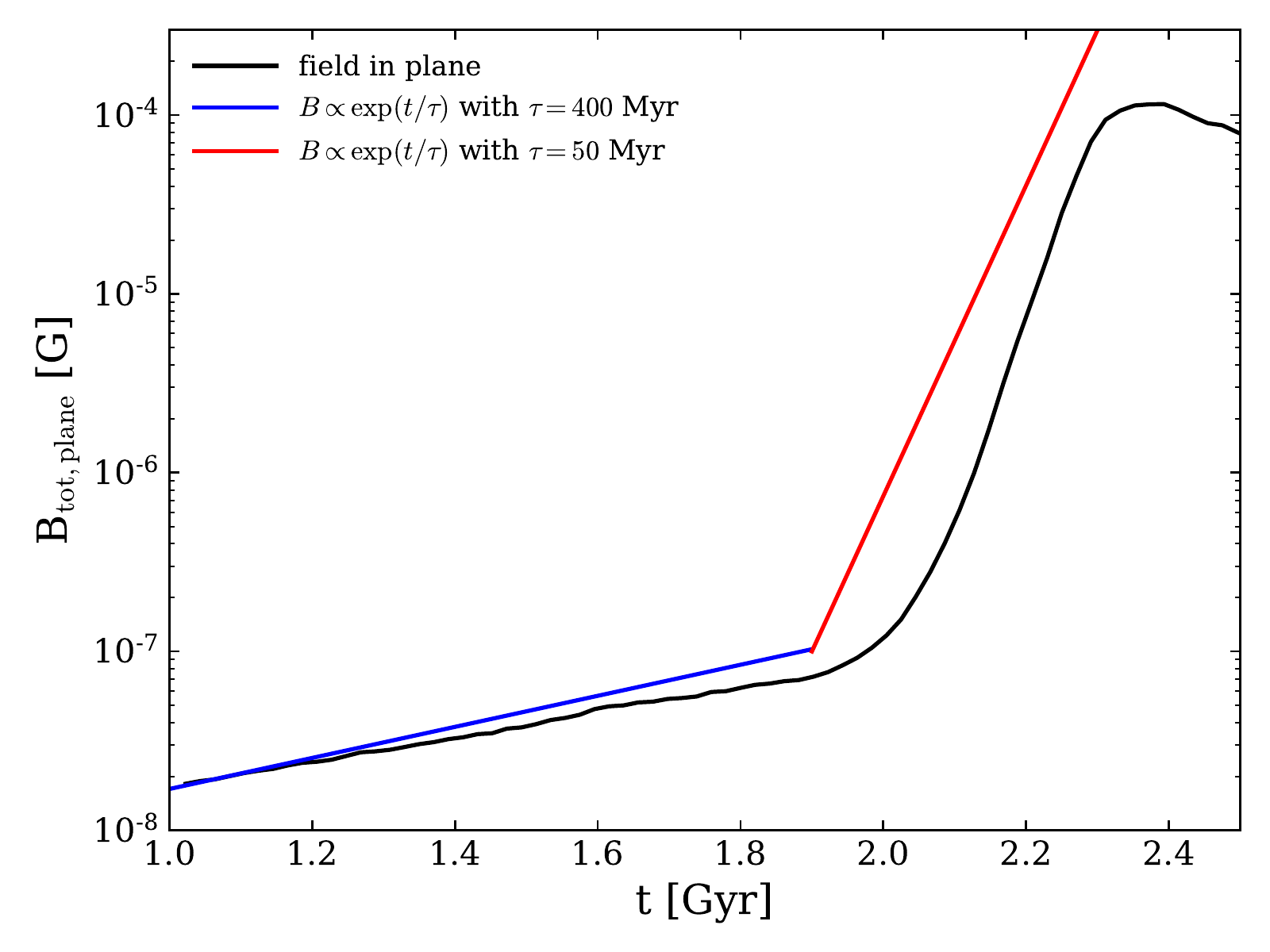}
    \caption{We show the magnetic field growth-rate in the innermost 2 kpc of the galaxy within a disc scale height of $200$ pc in the time frame of 1.0 Gyr to 2.5 Gyr (top). The field grows exponentially until 1.9 Gyr via a small-scale-turbulent dynamo with an eddy-turnover time of around $400$ Myr. After 1.9 Gyr the dynamo growth rate increases by a factor of $5$ and the dynamo grows on a timescale of 50 Myr in the very centre. This is faster than the growth-rate expected from the small-scale turbulent dynamo alone. The strong increase in the growth rate seems to be correlated with the formation of a bar in the centre and the increase of the field strength is roughly consistent with the prediction from the simple radial inflow dynamo. At 2.25 Gyr the growth is saturated by the large-scale outflow that is driven out of the central region with a low mass loading factor.}
    \label{fig:growth}
\end{figure}
\begin{figure}
    \includegraphics[scale=0.5]{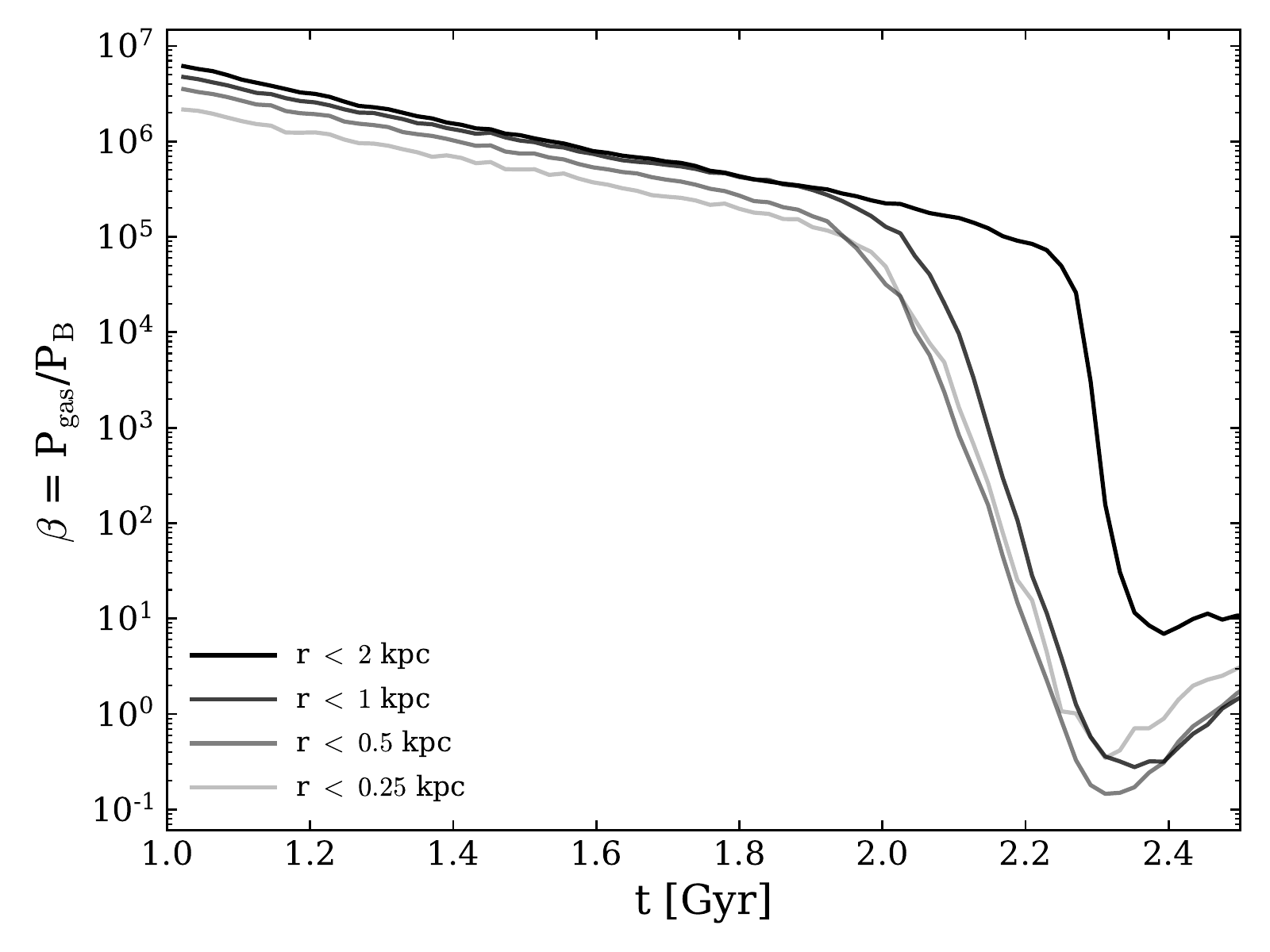}
    \caption{We show the plasma-$\beta$ for four different cuts for the innermost radius from $2$ kpc to $250$ pc. A value of $\beta > 1$ indicates that the fluid behaves hydrodynamical and the fluid is dominated by thermal pressure, while a value smaller than $\beta < 1$ marks the transition of a fluid that is dominated by the pressure provided by the magnetic field. As the magnetic field is amplified the center of the galactic disc transits from a state of thermal to magnetic dominated. Once the magnetic pressure dominates in the center, the outflow can be launched.}
    \label{fig:plasma_time}
\end{figure}
\begin{figure}
    \includegraphics[scale=0.5]{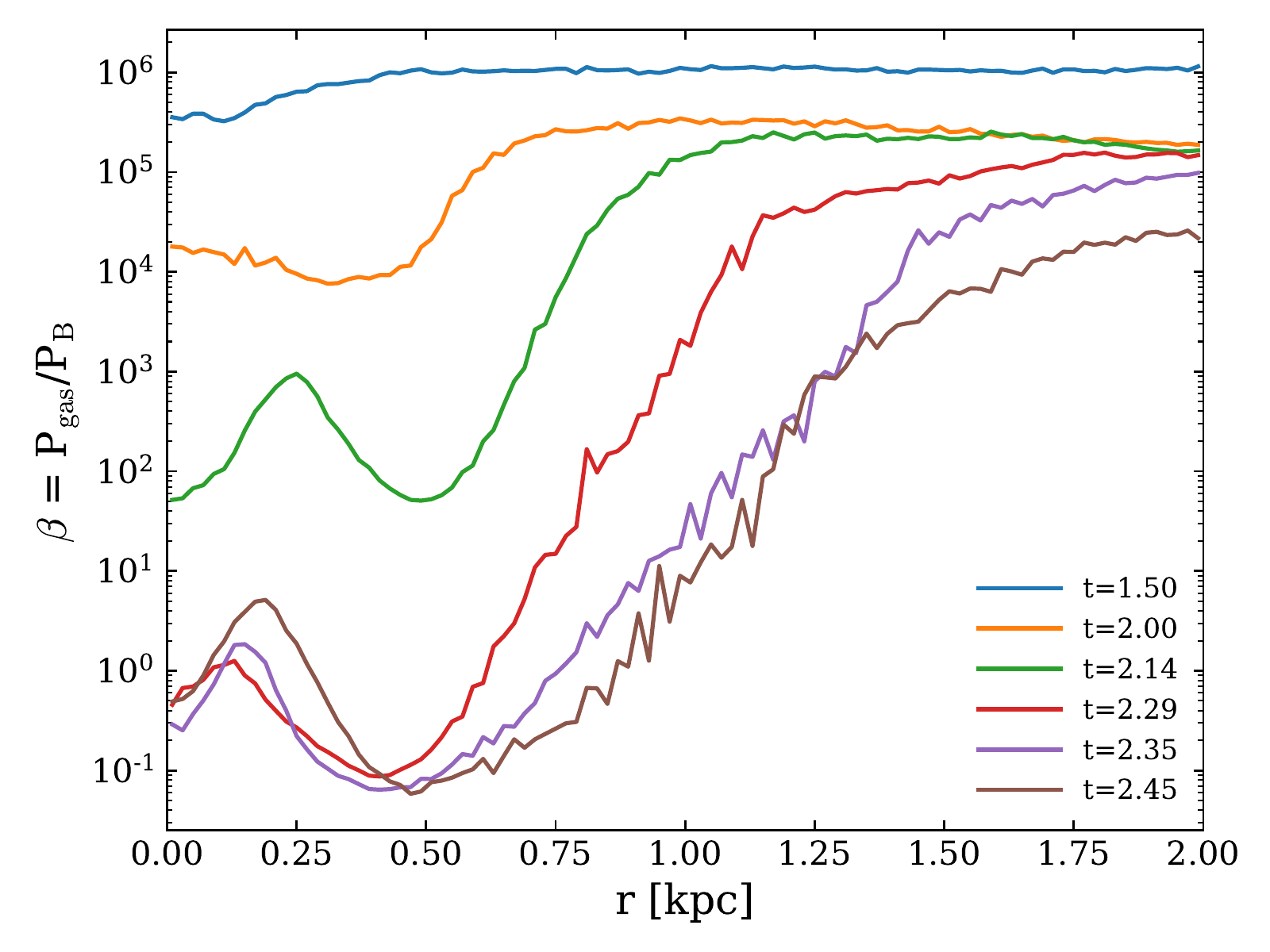}
    \caption{We show the radial-evolution of the plasma-$\beta$ for six different snapshots close to the outflow launching process. We see a strong decline of $\beta$ with time indicating the dominance of the magnetic field int he centre of the galaxy shortly before the outflow is triggered. This trend clearly shows that the outflow is subsequently launched because the magnetic field is much stronger than the thermal component.}
    \label{fig:plasma_radius}
\end{figure}

\begin{figure*}
    \centering
    \includegraphics[scale=1.0]{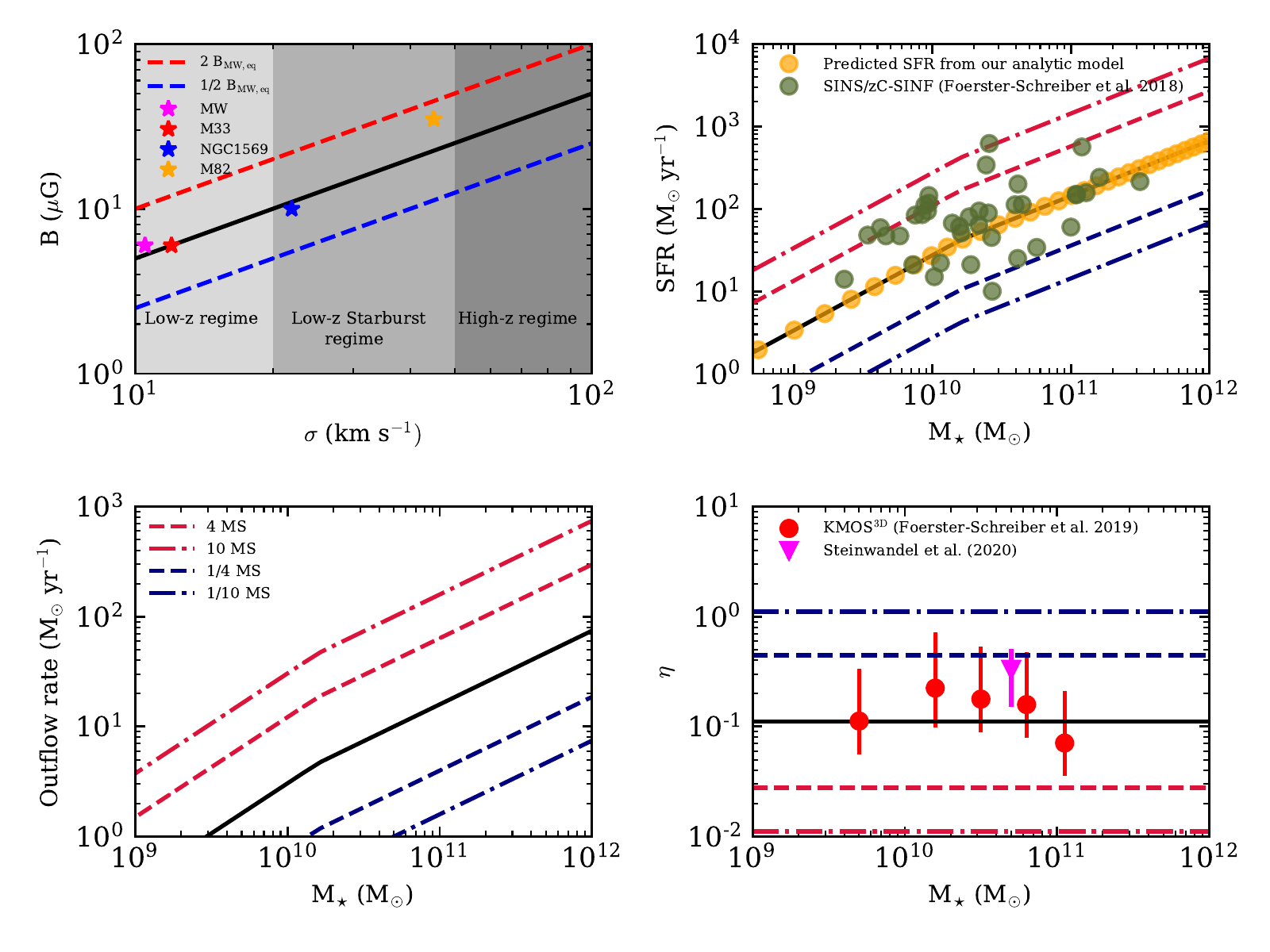}
    \caption{This Figure shows the main results of our simple toy model prescription for predicting SFRs, outflow rates and mass loading factors at $z \sim 2$. We assume that the magnetic field strength in equipartition scales with the turbulent velocity dispersion of the galaxy at hand, which would imply magnetic fields that are of the order of $30$ to $50$ $\mu$G in high redshift galaxies that typically have velocity dispersion of order $40$ to $120$ km s$^{-1}$ (top left). Moreover, we can compare this to a few galaxies where both, the magnetic field strength and the velocity dispersion can be observed (e.g. in the MW, M$33$, NGC$1569$ and M$82$). Via the magnetic field-star formation correlation of \citet{Schleicher2013} we obtain the SFR that corresponds to the higher magnetic field strength and cross-match this against the star formation main sequence at $z \sim 2$ taken from \citet{Withacker2014}. Finally, we can derive the outflow rate based on the theory of magneto-centrifugal winds. We find a power law increase of the outflow rate with stellar mass (bottom left). Finally, we can obtain the mass loading factor $\eta$ by dividing the outflow rate by the SFR (bottom right), which seems to be in good agreement with the mass-loading, that is obtained out of our simulation and the observations of the \textsc{kmos}$^{3\mathrm{D}}$-instrument. Furthermore the mass loading by such a magnetic driven wind appears to be constant which is for example different for cosmic-ray driven winds \citep[e.g.][]{Jacob2018} and could be used to distinguish between the two processes.}
    \label{fig:analytic_model}
\end{figure*}

\subsection{How to locally generate a super-equipartition field \label{sec:inflow_dynamo}}

There is some consensus in the literature that magnetic fields are amplified via such a process. The beauty of this is that the fast small-scale turbulent dynamo can quickly amplify the magnetic field, while the large-scale dynamo can order and retain it against magnetic diffusion for example due to reconnection events. However, there is an intrinsic problem with driving outflows based on the magnetic field in galaxies. First, magnetic fields are observed to be of the order of a few $\mu$G, which typically corresponds to some equipartition value with cosmic rays and often with the thermal pressure component as well. Driving an outflow via one of the non-thermal components becomes possible if it dominates the thermal component. In other words, for the magnetic field, the following condition has to be valid:
\begin{align}
    \beta = \frac{P_\mathrm{th}}{P_\mathrm{B}} = \frac{8 \pi \rho k_\mathrm{B} T}{B^2} \leq 1.0
\end{align}
where $\beta$ is the plasma parameter, based on the thermal pressure P$_\mathrm{th}$ and the magnetic pressure P$_\mathrm{B}$ of the fluid. At equipartition field strength this is not the case and the fluid settles at some low value of $\beta$ that is larger than one. However, to drive an outflow $\beta$ needs to be significantly smaller than one. Therefore, the first condition that is needed to drive an outflow via magnetic fields is to generate a super-equipartition field strength. We propose that this super-equipartition regime can be established by non-axis symmetric instabilities like bars, spiral arms or warps in galactic discs. Every non-axis-symmetric instability transports mass inwards and angular momentum outwards. In the specific case of a bar this leads to a gas response that is quicker than the outside co-rotation of the bar like mode with the rotation frequency of the bar $\Omega_\mathrm{p}$ equal to the rotation frequency of the galaxy $\Omega_\mathrm{g}$.\\ 
Under this assumption one can derive an upper limit for the growth of the magnetic field in toroidal direction by assuming that the flow is orientated alongside the bar and is of low velocity compared to the rotation of the bar. We can now derive the magnetic field evolution via the induction equation:
\begin{align}
    \frac{\partial \mathbf{B}}{\partial t} =  \nabla \times (\mathbf{v} \times \mathbf{B}),
    \label{eq:diff_without}
\end{align}
which gives us for the toroidal field component in thin disc approximation: 
\begin{align}
    \frac{\partial B_{\varphi}}{\partial t} = \frac{\partial v_{r} B_{\varphi}}{\partial r} + B_{r} r \frac{d\Omega}{dr}.
    \label{eq:radial_inflow}
\end{align}
We directly drop the diffusion term which will allow us to gain an upper limit on the magnetic field amplification via a rotational instability. However, we note that this introduces a number of problems as the diffusion term is for example needed to retain dynamo action within the plasma. Furthermore, by dropping the diffusion-term in equation \ref{eq:diff_without} we ignore the non-linear behaviour that will eventually lead to saturation of the field at a lower value. Calculations with the diffusion term have been carried out by \citet{Moss2000} who found a lower saturation value by $20$ to $30$ per cent accompanied by a slower growth rate of around the same order. We will discuss the effect of the diffusion term on our calculation in Appendix \ref{appendix:A} but note that we want to estimate an upper limit to estimate the significance of the effect.\\
For a fully developed bar like mode we assume that the magnetic field lines are already perfectly aligned with the bar. This only allows mass flux alongside the radial direction because the mass flux alongside the field lines is force free in ideal MHD. Thus $B_{r}$ remains roughly constant as the bar transports angular momentum outwards and mass inwards. We can then solve equation \ref{eq:radial_inflow} by integrating for fixed $B_{r}$ and we find \citep[e.g.][]{Lesch1993, Chiba1994}:
\begin{align}
    B_{\varphi} = \left[B_{\varphi, 0}(r_{0}) +\tau B_{r} r \frac{d\Omega}{dr}\right] e^{t/\tau} - \tau B_{r} r \frac{d\Omega}{dr}, 
    \label{eq:b_growth}
\end{align}
with the amplification time scale $\tau = -(\partial v_{r}/ \partial r)^{-1}$. Typically, the time-scale of such a process is of order of $0.1$ Gyr. If we now assume that the bar formation process takes $0.5$ Gyr, we obtain magnetic field growth of the toroidal component by factor of $\sim 150$ in the centre of the galaxy. If we assume a typical field strength of an already saturated field between $1$ $\mu$G and $10$ $\mu$G, obtained by the $\alpha^2$-$\Omega$-dynamo we obtain a central toroidal field between $100$ $\mu$G and $1000$ $\mu$G, which is in accordance with observations of the galactic centre \citep[e.g.][]{Yusef-Zadeh1996}. 

\subsection{Driving outflows with magnetic fields}

The final goal is now to derive an outflow rate that can be achieved by a magnetic outflow when the fluid has achieved super-equipartition field strength by a factor of $10-100$ of the equipartion field in the regime of a few $\mu$G. A very simple estimate can be obtained by making use of the two-dimdenional estimate we use for deriving the magnetic field growth. As we pointed out in section \ref{sec:alpha} the thin disc approximation has to fulfil the boundary conditions of B$_{r}$=B$_{\varphi} = 0$ at the edges of the disc.\\
While this is a problematic issue in the prescription of the $\alpha$-$\Omega$-dynamo as it would imply too large outflow rates and wrong energetics to actually obey this boundary condition, we can use it to calculate the outflow rate that one obtains due to the radial inflow dynamo. If we assume that the field grows by a factor of $100$, we can directly calculate the mass that moves out of the disc towards to CGM. The trick in this prescription is that we assume that the outflow is only triggered from the very centre and thus only a region with a mass of around $10^{9}$ M$_{\odot}$ is responsible for outflow action in a Milky Way-like galaxy. One can then assume that the disc had a mass of $M_{\mathrm{disc}, 1}$ before the radial inflow dynamo process started and the mass $M_{\mathrm{disc}, 2}$ after the process is suppressed again and following \citet{Kulsrud2008}:
\begin{align}
    M_{\mathrm{disc}, 2} = M_{\mathrm{disc}, 1} \cdot \left(\frac{B_{2}}{B_{1}}\right)^{-1/3}. 
    \label{eq:kulsrud_scaling}
\end{align}
The crucial assumption for this is that stellar feedback couples the magnetic field to the outflow. By assuming such a scaling we find that the central region can loose up to a fifth of its total mass due to this process, which results in an outflow rate of the order of $0.1$ M$_{\odot}$ yr$^{-1}$. Qualitatively, this is the easiest estimate for the outflow rate, but it is also problematic for two reasons. The major one is that such an outflow rate is enforced by the mathematical boundary conditions of how one treats the two dimensional MHD equations which is often used as an argument against the large-scale dynamo as the only mechanism that is held responsible for magnetic fields amplification. The second is, that it intrinsically depends on the chosen scaling in equation \ref{eq:kulsrud_scaling} and has no information of the exact coupling process of the magnetic field to the outflow. However, it can be used as an zeroth order estimate. \\
A better estimate can be obtained by considering the magneto-centrifugal outflow theory developed by several authors in regimes of stellar winds \citep{Weber1967, Mestel1968}, jets \citep{Blandford1982} and proto-stellar objects \citep[e.g.][]{Pudritz1983,Pelletier1992, Wardle1993, Shu1994, Spruit1996} which has also been applied to constrain the outflows in starburst galaxies \citep{DalPino1999} from a small disc around a central star cluster. The idea is that a collimated wind can be generated by a strong magnetic field in a disc-like configuration that allows for mass accretion. Typically, it is assumed that the magnetic field in the disc is generated by flux capture of the accreted material. \\
In our scenario the magnetic field growth is triggered by radial flows that enhance growth in the toroidial component. Obviously, in reality the magnetic field growth is much more complex than amplification via dynamos or radial flux capture from accreted material and is most likely a combination of those processes and also cosmic rays that compete with turbulent diffusion which will dissipate magnetic fields on the smallest scales.\\
Furthermore, we assume that the whole central disc of a Milky Way-like galaxy undergoes gravitational collapse due the formation of the bar that lead to magnetic field amplification which will subsequently drive the outflow. Thus, the major difference between the scenario of \citet{DalPino1999} and ours is that we start from a configuration of a stable disc that undergoes collapse due to non-axis-symmetric motions on the scale of roughly $1-2 $kpc, while \cite{DalPino1999} is investigating the outflow on the scale of a few $10$ to $100$ pc where a central nuclear disc forms around an active star cluster and the interaction of those two is driving the outflow. We further assume that the baryon over density increases towards the centres of galaxies and thus to first order, the same is valid for the energy densities of different components of the ISM as well (e.g. magnetic fields). \citet{Spruit1995} point out that such a configuration can lead to an opening field geometry with $\beta < 1.0$ which makes this scenario very interesting for our scenario at hand for explaining magnetic driven outflows in massive spiral galaxies. This is supported by the fact that the gas density right above and galactic disc can be orders of magnitudes lower than the disc material which further reduces $\beta$ just above the disc. \\
The field lines anchored in the disc can then support a flow from the disc towards the CGM and collimate it alongside field line perpendicular to the disc and control the opening angle of the outflow. We suggest that the opening field configuration can be supported by the Parker-Instability that can generate field lines perpendicular to the disc due to buoyant unstable flows under gravity. Thus our proposed wind scenario can straight forward establish as strongly collimated outflow. This is a fundamental difference between purely star burst driven winds which are weakly collimated. On top, the proposed scenario can explain the highly magnetised material driven outwards in superwinds in nearby galaxies like M82 \citep[e.g.][]{Baiaro2015, Rich2010, Roussel2010}. \\
In addition to $\beta < 1.0$ there are two crucial conditions for outflow launching via the magnetic field, a good coupling between the neutral component and the ions and a high magnetic Reynolds number R$_{m}$. This is both the case as one can carry out a similar estimate as \citet{DalPino1999} for a disc that is a factor of $10$ larger and sits in the center of a Milky Way-like galaxy, but yields a similar coupling constant between ions and neutral gas in a similar high magnetic Reynolds number flow. Thus we have conditions like presented in \citet{DalPino1999} and can make a similar estimate for the outflow rate. The key for the success of such a model is a good understanding for the accretion rate towards the galactic center. \\
The idea behind this is that there is tight coupling between the angular momentum in the central galactic disc and the angular momentum of the wind. This has an interesting consequence as the wind can then easily remove the angular momentum from the center of the galaxy and support the gravitational collapse of gas towards the galactic center. In classic magneto-centrifugal wind models one can then find an outflow rate via:
\begin{align}
\dot{M}_\mathrm{w} = f \cdot \dot{M}_\mathrm{a},  
\end{align}
where $\dot{M}_\mathrm{w}$ is the wind mass loss rate and $\dot{M}_\mathrm{a}$ is the accretion rate towards the center. The parameter $f$ describes how much of the accreted mass to the center is coupled to the outflow and scales as $f=(r/r_\mathrm{A})^2$ where r$_\mathrm{A}$ is the Alfv\'en-radius. The Alfv\'en-radius is the radius in the galactic disc where the magnetic energy density is exceeding the turbulent velocity. We are mainly interested at the value of $f$ at the driving scale of the wind. Thus in order to estimate $f$ we need to estimate r$_\mathrm{A}$ or directly the ratio of the driving scale to r$_\mathrm{A}$ as for example done in \citet{DalPino1999}, who estimate $f \approx 0.1$, which is in good agreement with the general understanding of magneto-centrifugal wind theory.\\
For now we assume that the driving scale of the wind corresponds to a length-scale of around $250-500$ pc and assume that the Alfv\'en-radius is larger by a factor of two to three. Furthermore, we assume that a bar within the Milky-Way can transport between $1-2$ M$_{\odot}$ yr$^{-1}$ towards the galactic center and a star formation rate between $0.5-1.0$ M$_{\odot}$ yr$^{-1}$ in the region of interest due to the mass inflow over the bar. These values are quite moderate and will lead in combination to a factor $f=0.1$, which seems to be in good agreement with other astrophysical systems for which outflow rates have been calculated via magneto-centrifugal wind theory. \citep[e.g.][]{Pelletier1992}. \\
From this we can obtain an outflow rate by this process in a Milky Way-like galaxy that corresponds to $0.05-0.4$ M$_{\odot}$ yr$^{-1}$. Considering the star formation assumed above this results in low mass loading factors $\eta$ with a numerical value of $eta$ around $0.1$. If such a process would act in the central region of the Milky Way on top of the mass loss due to AGN and supernovae the central region could loose up to $2 \cdot 10^{8}$ M$_{\odot}$ over a time scale of $500$ Myr. For a galaxy like the Milky Way this would imply that a fifth of the mass of the central region could be ejected towards the CGM via this process.\\
We note that the values we assumed so far appear to be quite arbitrary but as we will see in the next section they are in good agreement with the parameters for the Alfv\'en-radius, the driving-scale of the wind and the mass accretion rate over the bar and the star formation rate in the center of our numerical simulation of a Milky Way-like galaxy.

\section{Evidence from numerical simulations and the situation at \texorpdfstring{$z \sim 0$}{sec:sims}}
%\section{Evidence from numerical simulations and the situation at $z\sim 0$} 
\label{sec:sims}

Considering numerical simulations there is evidence on galaxy cluster \citep[e.g.][]{Dolag2002, Vazza2018, Rho2019} and galaxy scales \citep[e.g.][]{Butsky2017, Rieder2016, Rieder2017a, Rieder2017b, Pakmor2017, Martin2018, Steinwandel2019} for a small-scale turbulent dynamo that can at least be quantified over the Kazantsev-spectrum via an inverse energy-cascade from small to large scales. For our model we describe the details of the small-scale turbulent dynamo in \citet{Steinwandel2019} and for the large-scale dynamo in \citet{Steinwandel2020} where we furthermore discuss the role of magnetic fields in driving galaxy outflows based on excess of magnetic pressure over the thermal gas pressure-background of the galaxy. In the following we will develop a simple toy model that can be used to obtain outflow rates and mass loading factors by a wind that is initialised by the magnetic pressure alone, informed by our full three dimensional multi-physics simulation of a Milky Way-like galaxy.\\
Analogous to our estimates from section \ref{sec:inflow_dynamo} the outflow in our simulation is initialised by the formation of a bar. We show a volume rendering of our simulation of the gas density and the magnetic field in Figure \ref{fig:volume} shortly after the formation of the outflow at $t=2.233$ Gyr. In Figure \ref{fig:bars} we show the bar formation process in the inner region of the galaxy, which leads to excess mass inflow to the center of the galaxy. Mass can only flow parallel to the field lines in this configuration. Thus, angular momentum is transported outwards and mass inwards over the bar. This is the exact setup that we describe in section \ref{sec:inflow_dynamo}. We note that the central region keeps accreting mass over the bar once the outflow is launched as the outflow can effieciently transport angular momentum out of the center. In the simulation we find a radial inflow velocity of $\sim 1$ km s$^{-1}$ which is enough to trigger significant magnetic field growth in toroidal direction via equation \ref{eq:b_growth}. We gauge this in Figure \ref{fig:growth} where we show the growth of the magnetic field in the innermost $2$ kpc of the galaxy which is the region from where the outflow is launched. First the magnetic field is amplified from a zero background field via the small-scale turbulent dynamo in the center of the galaxy, which is indicated by the blue line which corresponds to the typical eddy-turnover time fro the small-scale turbulent dynamo in the ISM. At roughly $1.9$ Gyr of evolution of the galactic disc we find that there is steep increase of the growth rate by around a factor of for which shortens the characteristic implication time of the magnetic field to $50$ Myr. This is faster than typical growth rates of the small-scale turbulent dynamo and is the exact point where we can identify the bar formation process in Figure \ref{fig:bars}. Furthermore, the growth-rate of this process that we find in the numerical simulations is roughly consistent with the growth rate that we can obtain from equation \ref{eq:b_growth}. As our argument is based on the innermost $2$ kpc of our simulation we need to show that the magnetic field has a physical origin and is not amplified by numerical errors. We discussed this already in depth for the simulation at hand in \citet{Steinwandel2019} where we discussed that the numerical divergence is mostly problematic in regions with sharp density gradients, like the transition regions from spiral to inter-arm regions. Furthermore, we looked into the behaviour of the divergence of the central region and can confirm that it is decreasing as a function of time which renders it sub-dominant for amplification of the magnetic field in our simulation.\\
The crucial condition for launching an outflow from any component of the ISM is the pressure dominance of the specific component. Classic feedback processes like stellar feedback in the form of winds, radiation and supernovae from massive stars launch outflows by increasing the ISM midplane thermal pressure \citep[e.g.][]{Kim2015, Hu2016, Hu2017, Hu2019, Steinwandel2020b} and the subsequent formation of superbubbles. In this context the magnetic pressure has to be of leading order. In other words the plasma parameter $\beta = P_\mathrm{gas}/P_\mathrm{B}$ yields $\beta < 1.0$. \\
In Figure \ref{fig:plasma_time} we show the time evolution of the plasma parameter $\beta$ in the center of the galaxy for four different cuts for the radius between $2$ kpc and $0.25$ kpc. The galactic center transits from thermal pressure support towards magnetic pressure support, starting at $t=1.9$ Gyr with the onset of the bar formation process in the galaxy, indirectly confirming the strong magnetic field growth initialised by the bar. \\
It is interesting to point out that the region between $0.5$ kpc and $1.0$ kpc is most dominant in establishing $\beta < 1.0$. We further note that the outflow is not launched until the innermost region transits to $\beta < 1.0$. We show further evidence for this in Figure \ref{fig:plasma_radius} where we show the radial evolution of $\beta$ in the innermost $2$ kpc for six different points in time. Early into the evolution of the galaxy the center is completely dominated by the thermal gas pressure. Once the bar formation starts this is quickly changing and the central region is dominated by magnetohydrodynamical behaviour rather than hydrodynamical forces.\\ However, the issue with every outflow process is the coupling of the energy that is stored in the pressure to the ambient medium. In the case of common thermal feedback processes this happens by thermal gas heating and subsequent thermalisation of the hot component to kinetic energy. This question is a somewhat more tedious to answer in the case of magneto-centrifugal outflow. In \citet{Steinwandel2020} we pointed out that there is some evidence that the wind mechanism in the simulation is supported by the Parker-instability, which can be identified over the classic Parker-like lobes in the structure of the magnetic field lines that are lifted up from the central part of the disc and expand into the lower density CGM. Thus the Parker-Instability could account for the field geometry necessary for launching the outflow.\\
Moreover, the Parker-like lobes could then directly account for the common X-shaped halo field that is observed in many galaxies that are classified as out-flowing in the \textsc{chang-es} sample of nearby spiral galaxies. If we calculate the outflow rate for our Milky Way-like model using the prediction of magneto-centrifugal theory we find outflow rates of the process of the order of $0.05$ to $0.4$ M$_{\odot}$ yr$^{-1}$ which is in very good agreement with outflow rates that we find within our simulation that show values around $0.01-0.3$ M$_{\odot} yr^{-1}$ resulting in mass loading factors around $0.1$, as can be seen by the magenta point in the bottom left panel of Figure \ref{fig:analytic_model} alongside with the $2 \sigma$ percentiles on the error-bar. \\
In combination with the results of the \textsc{chang-es} collaboration who report outflow activity and an X-shaped halo field in a lot of the galaxies in their sample. We cross-correlated all the galaxies from their sample that could be classified as out-flowing with X-shaped halo field against their Hubble-type and find that at least $17$ of $22$ galaxies\footnote{We specifically refer to the galaxies NGC$891$, NGC$2820$, NGC$3003$, NGC$3044$, NGC$3079$, NGC$3432$, NGC$3556$, NGC$3735$, NGC$3877$, NGC$4013$, NGC$4157$, NGC$4217$, NGC$4302$, NGC$4565$, NGC$4666$, NGC$5775$ and UGC$10258$. Furthermore we note that for NGC 4565 the Spitzer Space Telescope revealed a bar in \citet{Barentine2009}, but it was classified as a grand design spiral before.} of the \textsc{chang-es} galaxies \citep[see][and references therein]{Krause2018, Krause2020} can be classified as barred-spiral galaxies. Whether or not our proposed process could play a role for outflows in galaxies could be tested with the upcoming Square Kilometre Array (SKA) in combination with the next generation of IFU-surveys that can constrain the kinematic information needed for identifying bars and other non-axisymmetric instabilities.\\
However, we used a very simple estimate for the ratio of the driving radius of the wind to the Alfv\'en-radius that we obtained from our simulation, which is in accordance with typically derived values for $f$ around $0.1$ for various physical systems \citep[see for different applications][]{Pelletier1992} and we further assumed typical star formation rates and inflow rates from our simulation using the results of our previous work \citep{Steinwandel2019, Steinwandel2020}. Specifically, the model should be improved for directly accounting the additional mass-accretion due to the bar on the side of the applied wind model. On the side of the numerical simulation we need better contrainss on teh driving scale and on the Alfv\'en-radius to develop a more conclusive mode in the future. While our simple estimates should be improved in the future as they only give an first order estimate of the outflow rate that are motivated by the findings of our simulation we find good agreement in terms of the mass loading factor via such a process.  
\\
We want to briefly discuss the consequences of such a magnetic driven outflow for cosmic-ray driven winds in galaxies. Recently, several groups revived the idea that cosmic-rays can significantly contribute to outflows in galaxies \citep[e.g.][]{Hanasz2013, Girichidis2016, Pakmor2016, Pfrommer2017, Hopkins2020c, Hopkins2020b, Hopkins2020d, Hopkins2020a} . This idea is quite intriguing due to the long cooling times of high energy cosmic-rays with respect to the lifetime of galaxies. The general idea of cosmic-ray driven winds is hereby to generate dominance of the cosmic-ray pressure over the thermal gas pressure either by cosmic-ray streaming or diffusion. While cosmic-ray streaming generates winds above the midplane, cosmic-ray diffusion can generate a wind at the base of the disc. Hence, diffusion driven outflows seem to be stronger as they can expel more gas from the disc. \\
For diffusion driven winds this intrinsically depends on the numerical value of the diffusion coefficient that accounts for the coupling. An outflow process as we presented it here could potentially be further enhanced by cosmic-ray driven winds, which will be subject of future work. 
Finally, we note the remarkable resemblance of the structural form of the outflow that we present in Figure \ref{fig:volume} with the ewly discovered structures above and below the midplane of the Milky Way with \textsc{erosita}. \citet{Predehl2020} showed that the structures that are typically referred to as the Fermi-bubbles extend much further out into the Milky Way halo. Our simulation indicates that a magnetic driven outflow could form these structures quite efficiently and we think the magnetic field of the Galaxy could play an important role in the formation of these structures alongside with the AGN in the galactic center.  

\section{Consequences for high redshift galaxies at\texorpdfstring{$z \sim 2$}{sec:highz}}
%\section{Consequences for high redshift galaxies at $z \sim 2$} 
\label{sec:highz}

Major results from high redshift observations show that the high redshift galaxy population is very compact and turbulent with thick discs, has strong galactic outflows and declining gas ration curves \citep[e.g.][]{Genzel2014, Genzel2018}.\\ 
There are some indicators in the line-of-sight velocity profiles \citep{Genzel2014} that these outflows are driven by the feedback of AGN or star-burst events. However, the line-of-sight velocity profiles indicate structure that allows us to speculate on other outflow mechanisms. We can use our derived outflow process from section \ref{sec:inflow_dynamo} and generalise it for the high redshift population to predict the impact of magnetic outflows in this environment. However, we note that our scenario is completely consistent with that of a star burst driven outflow where a star cluster is forming in the centre of the galaxy, that is surrounded by a disc that keeps accreeting mass and increases the magnetic field via flux capture which will generate a low $\beta$ environment needed for launching the outflow \citep{DalPino1999}.\\
On top of this it is unlikely that in such a regime as present, at $z \sim 2$, magnetic field amplification takes place via the $\alpha$-$\Omega$-dynamo as its timescale increases with $h^2$ and at $z \sim 2$ galaxies show thick discs with declining gas rotation, increasing the turbulent resistivity and decreasing the rotational support thus suppressing any $\alpha$-$\Omega$-dynamo action.\\ 
These systems are highly turbulent. The high amount of turbulence in discs at $z \sim 2$ can start magnetic field growth via the small-scale turbulent dynamo (kinematic regime) on Myr time scales and the magnetic energy density would quickly establish equipartition with the turbulent energy density, yielding $B \propto \sigma_\mathrm{turb}$ and increasing the magnetic field strength in high redshift systems easily by a factor of $5$.\\
From this we can directly estimate the SFR of these systems by applying the theoretical scaling of $B \propto \Sigma_\mathrm{sfr}^{1/3}$ that can be obtained analytically following \citet{Schleicher2013}, with the proper redshift correction. 
A non-axis symmetric instability like a bar, but also disc fragmentation \citep[e.g.][]{Behrendt2015} or cold filament accretion \citep[e.g.][]{Dekel2009} can now trigger magnetic field amplification via the radial motions. One can cross match the obtained SFRs of our model with the star formation main sequence (MS) at redshift $z \sim 2$ from which we obtain stellar masses which we can use to calculate the outflow rate by applying the theory of magneto centrifugal winds and use it in the regime at $z \sim 2$. \\
The later is the self-consistent way to derive the outflow rate. Therefore, we derive the outflow rate under the assumption that the star formation activity in high redshift galaxies comes from gas mass that was accreted to the galaxy. We further assume that the driving scale of the outflow is a factor of around three lower than the Alfv\'en-radius of the system, which comes from our redhsift zero simulation. This directly implies that $f=0.1$ is valid. This is a potential caveat and requires more simulations on our part to improve constraints on driving scales and Alfv\'en-radii as a function redshift. \\
Nevertheless, we can use this easy scaling to obtain an outflow rate and subsequently the mass loading of a magnetic driven wind for high redshift galaxies. This results in values below $1$, which is in agreement with the results from \citet{Foerster2018} for the \textsc{sins/zc-sinf ao} survey and from \citet{Foerster2019} with \textsc{kmos}$^{3\mathrm{D}}$. We show the results for this simple model in Figure \ref{fig:analytic_model}, where we show the relation between magnetic field and velocity dispersion (top left), our predicted SFRs (top right), outflow rates (bottom left) and mass loading's (bottom right) as function of stellar mass. \\
We note that this intrinsically depends on the shape of the star forming main sequence (MS) at the relevant redhshift, which is a clear limitation of the model which we plan to incorporate in future work. Nevertheless, the resulting outflow rates and mass loading factors are consistent with the theoretical expectations for an energy/entropy driven outflow and are consistent with the low observed mass loading factors from \citet{Foerster2019}. \\
It is interesting to point out an important issue regarding the observed low mass loading factor in observations and the reality of the high mass-loading factors in numerical simulations, which can reach values above unity even a t injection of the underlying feedback model. In cosmological simulations the mass loading factor is typically a free parameter, to constrain the galaxy population at some target redshift for example via the stellar-halo mass relation or the mass-metallicity relation. Therefore, a physical process with low mass-loading that can quench star formation and control the mass growth of galaxies is of potential interest also in a cosmological context. However, we are aware of the fact that the mass loading factor in cosmological simulations is the total mass loading factor while the one which is constraint from the observations of \citet{Foerster2019} is connected to the non-star forming gas as they observe in H-$\alpha$. Therefore, it is potentially possible that there is a lot of mass transport in the cold gas that is simply unaccounted for by current observations, which could justify the higher mass loading factors in cosmological simulations. In this case our proposed process would still contribute to the mass loading factor in the Warm-Ionised medium which is extremely important for the baryon cycle of galaxies. We discuss this issues in greater detail in Appendix \ref{appendix:B}. Our predictions could be tested by evaluating the magnetic field strengths in $z \sim 2$ galaxies with SKA in combination with high resolution IFU-spectrographs that can reveal the kinematic structure of these galaxies.
 
\section{Conclusions} \label{sec:conclusions}

We discussed the possibility and the consequences of magnetic driven outflows across redshift. We pointed out that spiral galaxies with strong magnetic field in the order of a few $\mu$G should be able to drive magnetic outflows with low mass loading factors if certain conditions are met. First, there mus be a process that can amplify the magnetic field to equipartition and provides the observed large-scale field structure. From satte-of-the-art numerical simulations there is an overwhelming evidence for the small-scale-turbulent dynamo as the main amplification process for the steady-state magnetic field. However, the observed magnetic field structure that is correlated on kpc scales seems to be consistent with the classic picture of the $\alpha$-$\Omega$-dynamo which is assumed to be too slow in amplifying the magnetic field on Gyr-timescales to observed values. Thus we argue for an $\alpha^2$-$\Omega$ dynamo in which the small-scale dynamo is amplifying the field and the large-scale dynamo is ordering and retaining the field against magnetic dissipation. \\
We showed that a steady-state-field can undergo fast exponential growth via radial flows if the galaxy forms a bar in its evolution. We find that such a process can amplify the field by at least an order of magnitude over the time-scale of around $500$ Myr, which is enough to generate a plasmabeta of around $0.1$ which is the perfect environment for launching a galactic wind by magnetic fields. We further assume that the Parker-Instability is providing the magnetic field structure needed for launching the outflow. The outflow process that we suggest is driven from the central region of massive spiral galaxies once they become bar-unstable and has driving scale of a few $100$ pc. The outflow rates that we are obtain are consistent within the framework of magneto centrifugal wind theory in terms of the predicted outflow-rate and show excellent agreement with our numerical simulation. Furthermore, the proposed structure resembles the structure of the Fermi-bubbles which have recently been observed to be much larger than originally expected. Thus our outflow mechanism could also partially play a role in explaining the new structure above and below the midplane of the Milky Way as revealed very recently by eRosita \citep{Predehl2020}. \\
Moreover, the combination of our model predictions and our numerical simulations can directly explain why so many galaxies of the \textsc{chang-es} sample that show a bar also show signs of outflows \citep[e.g.][]{Krause2018, Krause2020}.\\ 
Furthermore, it is possible to extend our modelling to galaxies at $z \sim 2$. The model is able to predict the observed SFRs at $z \sim 2$. Cross-correlating the obtained SFRs with the MS at $z \sim 2$ yields the observed high outflow rates and low mass loading factors that are observed with \textsc{kmos}$^{3\mathrm{D}}$ in \citet{Foerster2019}, who also find evidence for non-axis symmetric perturbations of the galaxies at $z \sim 2$ \citep{Foerster2019}. \\
We propose  that our suggested outflow process can contribute to the baryon budget of galaxies at low and high redshift. We believe that this can be tested with future IFU-surveys at high redhsift in combination with the capabilities of SKA that will provide us with magnetic field strength and kinematic information out to large redshifts. 
\\
\section*{Data Availability Statement}
The snapshot data that is used in the manuscript can be obtained on request from the first author of this manuscript.
The simulations presented in the manuscript are carried out with the developers version of our code \textsc{gadget-3}. The code version alongside with the initial conditions of the simulations and all relevant files to run the simulations will be provided by the first author on reasonable request.

\acknowledgments
UPS acknowledges valuable comments by Eirini Batziou, Ludwig Boess, Aura Obreja and Joseph O'Leary.
UPS, KD, HL and AB acknowledge support from the Deutsche Forschungsgemeinschaft (DFG, German Research Foundation) under Germanys Excellence Strategy - EXC-2094 - 390783311 from the DFG Cluster of Excellence 'ORIGINS'\\
UPS acknowledges funding by the Deutsche Forschungsgemeinschaft (DFG, German Research Foundation) with the project number MO 2979/1-1. \\
UPS acknowledges computing time granted by the Leibniz Rechenzentrum (LRZ) in Garching under the project number pn72bu and computing time granted by the c2pap-cluster in Garching under the project number pr27mi.

%% To help institutions obtain information on the effectiveness of their 
%% telescopes the AAS Journals has created a group of keywords for telescope 
%% facilities.
%
%% Following the acknowledgments section, use the following syntax and the
%% \facility{} or \facilities{} macros to list the keywords of facilities used 
%% in the research for the paper.  Each keyword is check against the master 
%% list during copy editing.  Individual instruments can be provided in 
%% parentheses, after the keyword, but they are not verified.

%% Similar to \facility{}, there is the optional \software command to allow 
%% authors a place to specify which programs were used during the creation of 
%% the manuscript. Authors should list each code and include either a
%% citation or url to the code inside ()s when available.

%% Appendix material should be preceded with a single \appendix command.
%% There should be a \section command for each appendix. Mark appendix
%% subsections with the same markup you use in the main body of the paper.

%% Each Appendix (indicated with \section) will be lettered A, B, C, etc.
%% The equation counter will reset when it encounters the \appendix
%% command and will number appendix equations (A1), (A2), etc. The
%% Figure and Table counter will not reset.

\appendix

\section{Radial inflows and magnetic diffusivity}
\label{appendix:A}
\begin{figure*}
    \centering
    \includegraphics[scale=1.0]{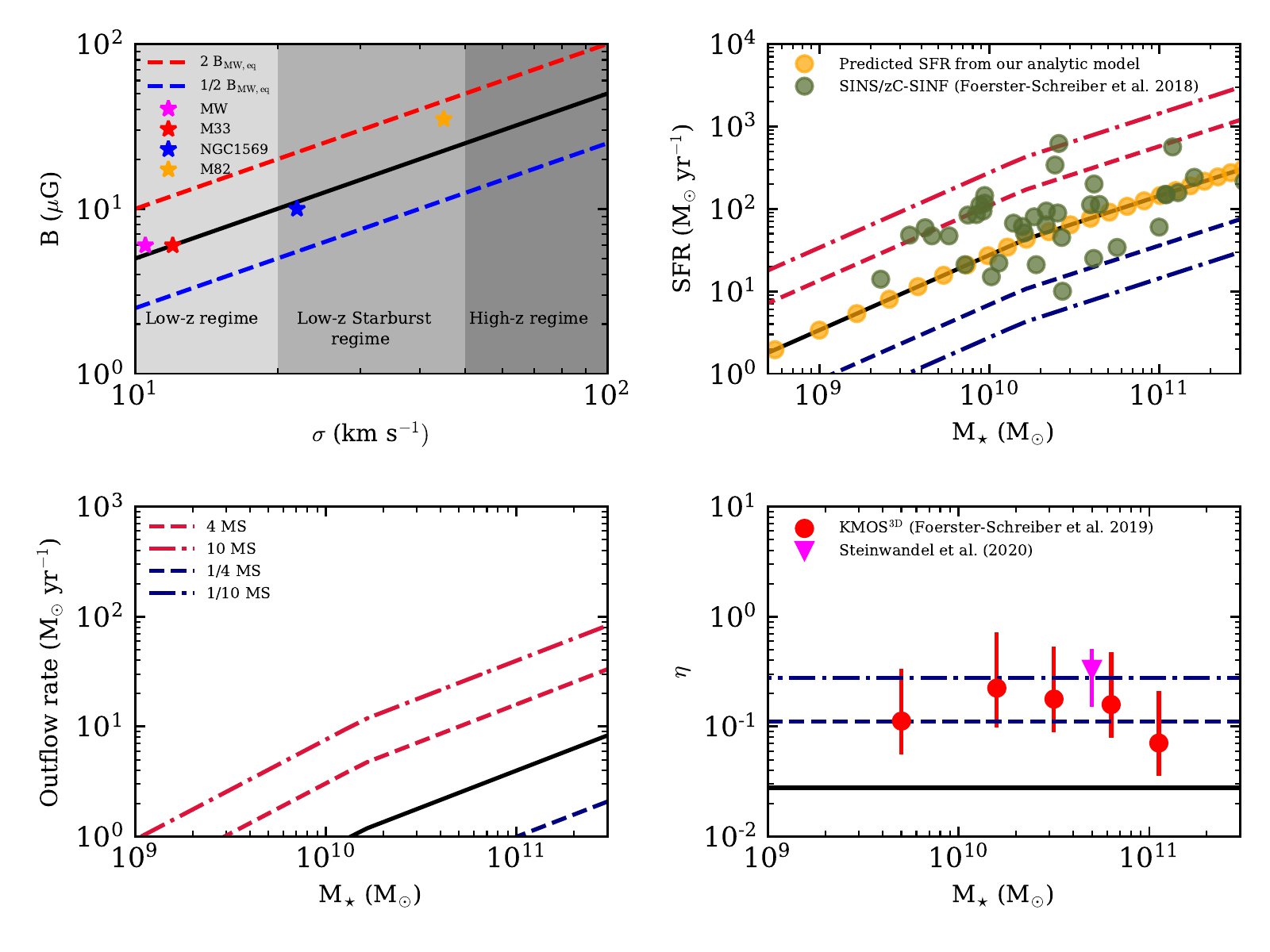}
    \caption{Same as Figure \ref{fig:analytic_model} but assuming that the magnetic field growth is suppressed by 30 per cent as pointed out by \citet{Moss2000}. However, we note that in our case the correction term of \citet{Moss2000} is quite small as we are closer to the center and the correction term scales with radius squared.}
    \label{fig:analytic_model_diff}
\end{figure*}

As we already pointed out in section \ref{sec:inflow_dynamo} neglecting the diffusion term leads to an upper limit of our modelling. The problem with dropping the diffusion term is two-fold. The first problem that arises is that the diffusion term is needed for a proper dynamo model. However, the idea of this paper is to get an easy analytic insight on how magnetic fields could potentially drive outflows over a wide range of galaxy mass and redshift and not to model the detailed influence of the radial inflow on the dynamo growth rate, which would be a different study. The second problem that arises by dropping the diffusion term is simply that there is no intrinsic dissipation of magnetic energy on the smallest scales, which leads to an overestimate of the field growth. This has already been pointed out before. \citet{Moss2000} solved the classic dynamo-equations under the assumption of an additional radial inflow of the order of 1 km s$^{-1}$ while accounting for the diffusion term. \citet{Moss2000} points out that \citet{Chiba1994} dropped and they render this as a strong oversimplification of the situation as \citet{Chiba1994} draw the conclusion from their result that radial inflows can explain magnetic field growth in galaxies altogether. While it is true that the model if \citet{Chiba1994} is missing the diffusion term we use this model in this work to get an analytic insight on the growth-rate of the magnetic field in the innermost $2 $kpc of our MW-like disc galaxy simulation of which we showed small-scale and large-scale dynamo growth in \citet{Steinwandel2019} and \citet{Steinwandel2020}. As \cite{Moss2000} brings forward the strongest criticism of the model of \citet{Chiba1994} we want to discuss why the regime that we consider is different from the regime that is dicussed in both, \citet{Chiba1994} and \citet{Moss2000}. The idea for our model is motivated by our simulations. As we showed in \ref{fig:growth} from our own simulations that there is evidence for a strong increase of the growth-rate of the magnetic field in the plane of the innermost $2$ kpc within the disc scale height of $200$ pc. The growth-rate of the magnetic field increases by a factor of $5$ when the bar formation process in the centre of the galaxy starts. The jump in the growth-rate and the bar formation are therefore tightly correlated which is clearly shown by our simulations and is in rough agreement with the predicted growth-rate for amplification by \citet{Chiba1994} within a factor of $2$. Our simulation indicates hereby slightly faster dynamo growth than the model from \citet{Chiba1994}. In contrast to this the detailed treatment and solution of the dynamo-equations by \citet{Moss2000} indicates that radial flows in barred-spiral galaxies could reduce the dynamo growth-rate by around $20$-$30$ per cent which is true for radial flows with positive and negative sign. However, a reduced growth-rate of the dynamo and a correlated lower saturation value do not indicate that there is not net growth of the field. \citet{Moss2000} back-up their results with an 2-dimensional simulation which is much different than our three-dimensional multi-physics simulation that resolves the adiabatic compression regime, the small-scale and the large-scale dynamo and incorporates a treatment for star formation and stellar feedback and is suited fro the investigation of magnetic field growth from a zero background field. Furthermore, the focus of \citet{Moss2000} and the argument on suppressed dynamo growth is focused on the conditions in the solar neighbourhood, which we do not apply as we strictly consider magnetic field growth within the innermost $2$ kpc where \citet{Moss2000} as an indication for driving outflows based on the magnetic pressure in local spiral galaxies. \citet{Moss2000} provides us with an estimate for the saturation value of the toroidal magnetic field at the solar radius for an radial flow of $1$ km s$^{-1}$ and find B$_{\varphi} \approx 10$ $\mu$G. However, if we assume more realistic values for our configuration we obtain a value of around B$_{\varphi} \approx 100$ $\mu$G for the saturation value which is in good agreement with our predicted saturation field strength from our simulations. This is related to the fact that the correction terms determined by \citet{Moss2000} are scaling with R$^{2}$ and v$_{r}^2$, where R is the distance from the centre of rotation and v$_{r}$ is the radial velocity. As the distance from the rotation centre is much smaller in our case we find a weaker deviation from the dynamo growth rate. Furthermore, we note that the situation is much more complicated in our case as we resolve the small-scale turbulent dynamo action. Thus, it could be possible that the amplification is only indirectly triggered by the inflow as star formation in the centre increases due to the radial inflow which will increase turbulence and might simply shorten the eddy-turnover time that is correlated with the growth-rate of the small-scale turbulent dynamo. However, as this amount of turbulence is triggered by the radial inflow the magnetic field amplification can then be estimated to first order by increase obtain over the radial inflow and the subsequent amplification of the magnetic field which appears to be consistent with the easy model by \citet{Chiba1994}. Consideration of the diffusion term would then lead to a slightly lower saturation value in the centre which would still be in agreement with our the saturation value for the central plane predicted by our simulations. We note that our argumentation slightly differs fro the high redshift systems as we intrinsically assume that feedback is so strong that those system constantly become gravitational unstable in their dark matter potential. We show the modified model predictions by taking a smaller growth rate of the dynamo in the centre into account in Figure \ref{fig:analytic_model_diff} from which we directly see that the growth-rate of the dynamo due to the radial-inflow is not dominating the mass loading, even if radial flows suppress the dynamo action by a 30 per cent margin.

\section{Deriving the outflow-rate by the dynamo boundary conditions}
\label{appendix:new}

\begin{figure*}
    \centering
    \includegraphics[scale=1.0]{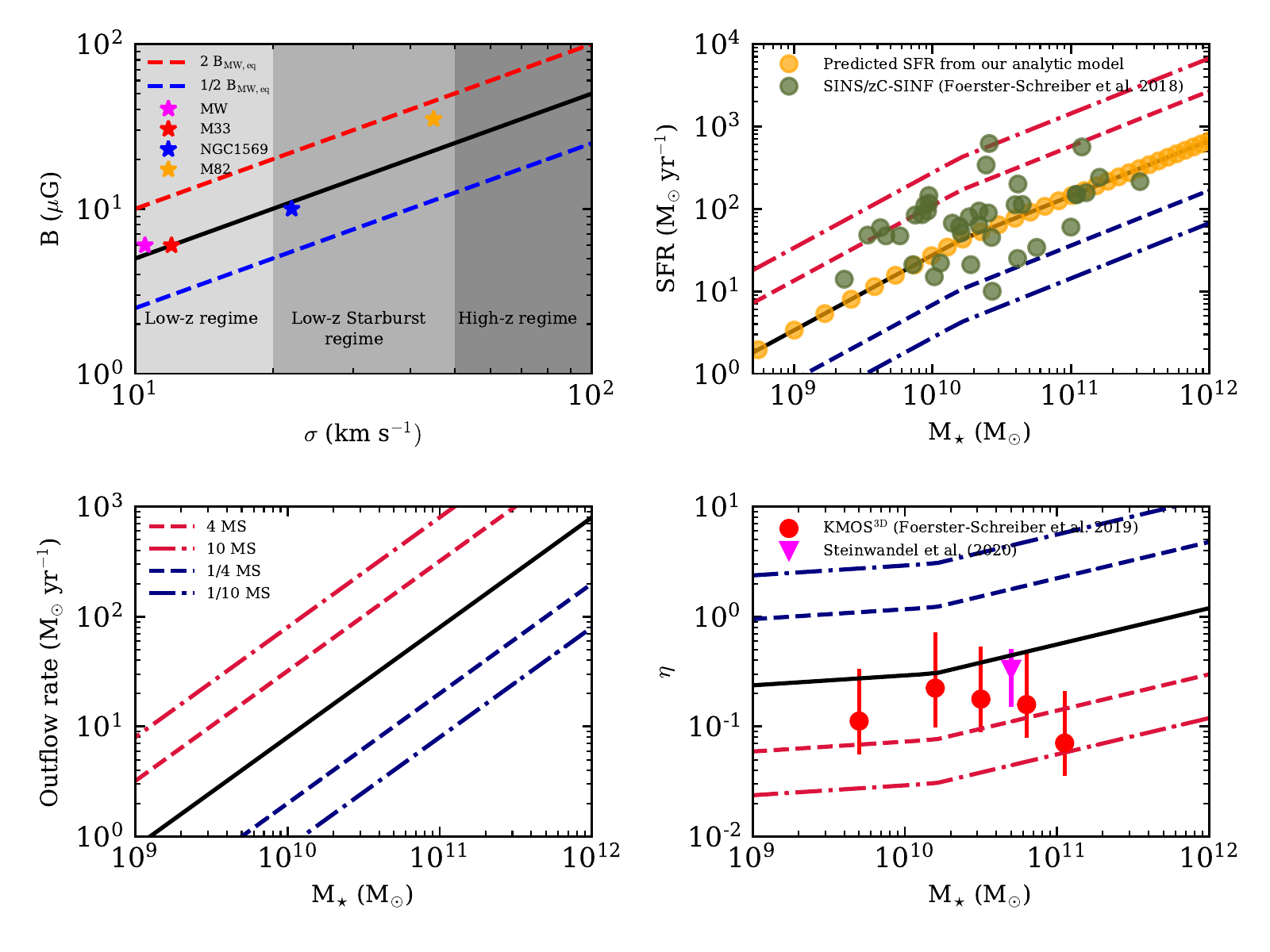}
    \caption{Same as Figure \ref{fig:analytic_model} but assuming the scaling from \citet{Kulsrud2008} and estimating the outflow properties by the boundary condition of vanishing magnetic field at infinity.}
    \label{fig:analytic_model_old}
\end{figure*}

We briefly discussed the possibility of deriving the outflow rate via the boundary conditions of a vanishing magnetic field at the edges of the galactic disc and how this would infer a mass transfer from the disc the CGM. In the classic picture of the $\alpha$-$\Omega$-dynamo this leads to an excess mass flux that would be not physical. However, in the case of our bar-instability outflow one could circumvent this problem by assuming that only the central region has to account for the outflow, which would result in realist outflow rates and mass loading factors around the same value that we can predict from magneto-centrifugal outflow theory. The biggest theoretical challenge by deriving the outflow rate in this fashion is that the mass flux arises to fulfil the unrealistic boundary condition at the discs edge that arise when one is solving the two-dimensional dynamo-equations. Moreover, in this picture it remains unclear how this to couple the outflow to the ambient ISM. It can be done by assuming that stellar feedback is providing the structures necessary to generate a vertical magnetic field. For completeness we show that this ansatz results in similar outflow rates than in magneto-centrifugal theory in Figure \ref{fig:analytic_model_old}. \\

\section{The interdisziplinary nature of this approach}
\label{appendix:B}
This work has a highly interdisziplinary nature and we want briefly put our work into the context of the different sub-fields that it is related to. First, the contribution in the area of galactic magnetic fields is apparent by the fact that our simulations show resolved dynamo action from the small-scale dynamo and the large-scale dynamo as we reported in \citet{Steinwandel2019} for the small-scale turbulent dynamo and in \citet{Steinwandel2020} for the large-scale dynamo. However, we find the leading order in magnetic field amplification is driven by the small-scale turbulent dynamo. Furthermore, our simulations are not the only ones that predict this outcome \citep[see e.g.][]{Rieder2016, Rieder2017a, Pakmor2017}. In our picture the large-scale dynamo is only needed for ordering the field on the larger-scales as the small-scale turbulent dynamo struggles to explain the large correlation lengths of the magnetic field in the Galaxy. However, \citet{Rieder2017a} suggest that the large-scale magnetic field structure could also be generated by in falling gas from the CGM. Furthermore, there is analytic work from \citet{Xu2020} which could explain the kpc correlation lengths due to the non-linear growth of the dynamo modes.  However, we find that the large-scale dynamo is the leading process to order the field on the larger scales to a quadrupolar structure which is consistent with observations \citep{Stein2019}.\\
Second, there is the outflow aspect of this work. Our simulations indicate an outflow with a low mass loading factor for MW-like spiral galaxies. In this context we point out the interesting aspect that observed local spiral galaxies that show a sign of outflows seem to be heavily biased towards being classified as barred spiral galaxies. Often these outflows are explained either by stellar feedback or by the cosmic-ray pressure component that can launch them from the ISM. We ask the simple question what if these outflows are driven by the magnetic pressure instead of the cosmic ray pressure due to strong magnetic field amplification in the central part of massive spiral galaxies that undergo a gravitational instability and drive magnetic field growth in the centre as a combination of adiabatic compression and a fast dynamo process. Our outflow due to the magnetic field structure provided by the buoyancy (Parker) instability in the magnetically over-pressurised medium, which is different from classic wind launching processes in galaxies. Our simplified model can either be confirmed or be out ruled with a combination of up-coming surveys like SKA and the next-generation of high resolution IFUs that can give an insight on the detailed gas structure.\\
Finally, there is the galaxy formation aspect of this work for which we try to evaluate the importance of magnetic fields in massive galaxies at higher redshift and discuss the consequences gravitational (bar-like) modes and elaborate if a strong magnetic field at high redshift could launch an outflow. by doing so we find that our very simple model that is easy to understand predicts magnetic driven outflows at higher redshift with a low mass-loading. Numerical simulations (specifically large cosmological volumes) typically assume some mass loading which is much higher than suggested by observations. This is specifically true for the galaxies at the high mass end of the stellar mass function. In large cosmological volumes the mass loading $\eta$ is typically a free parameter of the modelling which is tuned to reproduce some quantity at redshift zero (e.g. the stellar mass function or the mass metallicity relationship of galaxies). Our simulations suggest a process that can quench star formation by self-consistently establishing a magnetic outflow with very low mass-loading in agreement with observations at low redshift. In combination with our simple toy model we can investigate if such a process can establish as similarly outflow with low mass loading at higher redshift. Such processes are important to study because they are decoupled from the thermal feedback loop of galaxies and the observed variations in the velocity line profiles of high redshift galaxies would potentially allow for different feedback channels apart from thermal feedback by supernovae, stellar winds and AGN. 

%% For this sample we use BibTeX plus aasjournals.bst to generate the
%% the bibliography. The sample63.bib file was populated from ADS. To
%% get the citations to show in the compiled file do the following:
%%
%% pdflatex sample63.tex
%% bibtext sample63
%% pdflatex sample63.tex
%% pdflatex sample63.tex

\bibliography{sample63}{}
\bibliographystyle{aasjournal}

%% This command is needed to show the entire author+affiliation list when
%% the collaboration and author truncation commands are used.  It has to
%% go at the end of the manuscript.
%\allauthors

%% Include this line if you are using the \added, \replaced, \deleted
%% commands to see a summary list of all changes at the end of the article.
%\listofchanges

\end{document}